\documentclass[pra,showpacs,twocolumn]{revtex4}
\usepackage{bm,graphicx,amsmath}
\usepackage{epsfig}
\begin{document}

\title{Cold Fermi atomic gases in a
pumped optical resonator}
% \title{}
\author{Jonas Larson$^{1,2}$, Giovanna Morigi$^3$ and Maciej
Lewenstein$^{1,4}$} \affiliation{$^1$ICFO--Institut de Ci\`encies
Fot\`oniques, E-08860 Castelldefels, Spain}
\affiliation{$^2$NORDITA, 106 91 Stockholm,
Sweden} \affiliation{$^3$ Departament de F\'{i}sica, Universitat
Aut\`{o}noma de Barcelona, E-08193 Bellaterra, Spain}
\affiliation{$^4$ICREA-- Instituci\'o Catalana de Recerca i Estudis
Avan\c cats, E-08010 Barcelona, Spain}

%\address{ICFO-Institut de Ci\`{e}ncies Fot\`{o}niques, E-08860 Castelldefels, Barcelona, Spain}
\date{\today}

\begin{abstract} We study systems of fully polarized ultracold
atomic gases obeying Fermi statistics. The atomic transition
interacts dispersively with a mode of a standing-wave cavity, which
is coherently pumped by a laser. In this setup, the intensity of the
intracavity field is determined by the refractive index of the
atomic medium, and thus by the atomic density distribution.
Vice versa, the density distribution of the atom is determined by
the cavity field potential, whose depth is proportional to the
intracavity field amplitude. In this work we show that this
nonlinearity leads to an instability in the intracavity
intensity that differs substantially from dispersive optical
bistability, as this effect is already present in the regime, where
the atomic dipole is proportional to the cavity field. Such
instability is driven by the matter waves fluctuations and exhibits
a peculiar dependence on the fluctuations in the atomic density distribution.
 \end{abstract}

 \maketitle

\section{Introduction}

Cavity Quantum Electrodynamics (CQED) is one of the most rapidly
developing areas of modern quantum optics and atomic molecular, and optical physics~\cite{walls-book,berman-book}. In its early days optical
instabilities, with the paradigm example of optical bistability were
in the center of interest \cite{obexp1,obexp2}. Optical bistability
emerges from the nonlinear response of the dipolar transitions of
atoms to the cavity field, to which they couple. Theoretically, such
dynamics are well described by mean field theories formulated for
macroscopic variables, such as the cavity field intensity and the
total atomic polarization~\cite{carmichael,MW,narducci}.
Fluctuations of the polarization, in particular in the quantum
regime, may play an important role in these processes, as pointed
out for instance in the seminal studies of the late Dan Walls and
collaborators (see \cite{walls-book} and references therein).
Typically, however, these dynamics are characterized by few {\it
macroscopic} quantum degrees of freedom, where quantum fluctuations
are small, except at the instability points. Nonlinearity at the
microscopic level has since then been reached by high-finesse
resonators, where the strong coupling regime between a single atom
and a single photon have been realized in several milestones
experiments~\cite{walther-review,haroche-book,others}. In this
regime the quantum dynamics of few {\it microscopic} quantum degrees
of freedom dominates the behavior of the system, and quantum
fluctuations are typically large.

Recently experimental progress has allowed one to strongly couple
cold atomic gases with the electromagnetic field mode of a
resonator. This line of research, which we would like to term cavity
QED of many-body systems, was stimulated by the observation on the
interdependence between intracavity field and atomic motion via
of the mechanical effects of light~\cite{asboth,Domokos_JOSAB}. This was confirmed
by seminal experiments, which demonstrated cavity cooling of
atoms~\cite{Pinkse} and selforganization of atoms in regular
patterns inside pumped cavities~\cite{black,Collective}. Most
recently, the coherent coupling between a Bose-Einstein condensate
and a cavity field has been experimentally
demonstrated~\cite{ZimmermannBEC,Esslinger,Reichel}. Moreover, the
nonlinear interaction between ultracold atoms and the field of a cavity mode
had been experimentally observed \cite{stamper-kurn}. In
this new ultracold regime, theoretical work on bosonic atomic gases
showed that the matter-wave quantum fluctuations and the spatial
mode variance give rise to additional nonlinear atom-field
effects~\cite{Maschler}, leading to features of the phase diagram
which suggest novel bistability
phenomena~\cite{jonas1,jonas2,jonas3}.

In this work, we consider a gas of ultracold and polarized
fermionic atoms, whose dipolar transition interacts dispersively
with the mode of a standing-wave resonator. In the regime in which
the atomic polarization is linear in the field amplitude, and
hence where there is no usual optical bistability, we encounter
instabilities of the intracavity field which originate from the
coupling with the atomic motion. The theoretical model we apply is
based on a second-quantized approach for the description of the
atomic and of the field degrees of freedom, in the regime in which
the atomic internal degrees of freedom are eliminated from the
equations in second order perturbation theory. These dynamics are
studied by numerically integrating the Heisenberg equations of
motion. The results show that the instabilities of the intracavity
field intensity are driven by the pump strength and by the atomic
density, and indirectly depend on the degree of localization of
the atoms at the minima of the cavity potential.

This article is organized as follows. In Section~\ref{Sec:II} we
theoretically derive the model, in Sec.~\ref{Sec:III} we solve
numerically the coupled Heisenberg equations of motion for field
and atoms for a fixed number of atoms and thoroughly discuss the validity of the considered approximations and experimental parameters. The
conclusions and outlook are presented in Sec.~\ref{Sec:V}.

\section{Theoretical derivation of the model} \label{Sec:II}

In this section we derive the theoretical model, which is at the
basis of the studies presented in Sec.~\ref{Sec:III}. In order to
introduce the many-body theory, which is the starting point of our
study, let us first review the Hamiltonian of a single two-level
atom of mass $m$ interacting with a cavity field which is pumped
coherently by an external, monochromatic field. We denote by
$\hat{\sigma}=|g\rangle\langle e|$ and $\hat{\sigma}^\dagger$ the
atomic lowering and raising operators of the dipole transition at
frequency $\omega_a$ with ground and excited states
$|g\rangle,|e\rangle$, and by $\hat{a}$ and $\hat{a}^\dagger$ the
annihilation and creation operators of a cavity photon at energy
$\hbar\omega_c$. We further assume that the atomic motion is along
the cavity axis $X$, while the radial degrees of freedom are frozen
out, and denote by the operators $\hat{P}$ and $\hat{X}$ the atomic
centre-of-mass momentum and position. In the dipole and rotating
wave approximation, the Schr\"odinger equation of atomic and cavity
degrees of freedom is governed by the Hamiltonian \cite{jonas1}
\begin{equation}\label{jcham} \begin{array}{lll} H_{JC}\!
&\! =\! &\! \displaystyle{\frac{\hat{P}^2}{2m}\!-\!\hbar\tilde{\Delta}_a\hat{\sigma}^\dagger\hat{\sigma}\!-\!\hbar\tilde{\Delta}_c\hat{a}^\dagger\hat{a}\!-\!i\hbar g(\hat{X})\!\!\left(\!\hat{\sigma}^\dagger\hat{a}\!-\!\hat{a}^\dagger\hat{\sigma}\!\right)} \\ \\
& & -i\hbar\tilde{\eta}\left(\hat{a}-\hat{a}^\dagger\right).
\end{array} \end{equation} which is here given in the reference
frame rotating at the pump frequency $\omega_p$. The parameters
$\tilde{\Delta}_a=\omega_p-\omega_a$ and
$\tilde{\Delta}_c=\omega_p-\omega_c$ are the detunings of the pump
from atomic and cavity mode frequencies, respectively, and
$\tilde{\eta}$ is the pump amplitude. The dipole-cavity mode
coupling has position-dependent strength
$g(\hat{X})=g_0\cos(q\hat{X})$ with $q$ the cavity-mode wave number.
For large detuning $|\Delta_a|$ the internal atomic level can be
eliminated from the equations of cavity and atomic external degrees
of freedom~\cite{jonas2}. The effective Hamiltonian reads
\begin{equation}\label{jcdisp}
\begin{array}{lll}
H & = & \displaystyle{\frac{\hat{P}^2}{2m}+\hbar\tilde{U}_0\hat{a}^\dagger\hat{a}\cos^2(q\hat{X})-\hbar\tilde{\Delta}_c\hat{a}^\dagger\hat{a}} \\ \\
& & -i\hbar\tilde{\eta}\left(\hat{a}-\hat{a}^\dagger\right),
\end{array} \end{equation} with
$\tilde{U}_0=g_0^2/\tilde{\Delta}_a$, and where we have considered
temperatures for which the atomic motion is frozen
on the time scale in which the internal degrees of freedom
appreciably evolve.

For later convenience we introduce the dimensionless variables
\begin{equation} \begin{array}{lll}
\displaystyle{\hat{p}=\frac{\hat{P}}{\hbar q}}, &
\displaystyle{\hat{x}=q\hat{X}}, &
\displaystyle{t=\tilde{t}\omega_r,} \\ \\
\displaystyle{U_0=\frac{\tilde{U}_0}{\omega_r}}, &
\displaystyle{\Delta_c=\frac{\tilde{\Delta}_c}{\omega_r}}, &
\displaystyle{\eta=\frac{\tilde{\eta}}{\omega_r}.} \end{array}
\end{equation} where $\omega_r=\frac{\hbar q^2}{2m}$ is the recoil
frequency.

We remark that when neglecting the coupling between cavity and
motion, hence in the classical limit, Hamiltonian in
Eq.~(\ref{jcdisp}) does not give rise to instabilities like in
optical bistability. Dispersive optical bistability, in fact, would
appear when the next order in the perturbative expansion is
included, giving rise to a nonlinear (Kerr) term of the form
$\hat{a}^\dagger\hat{a}^\dagger\hat{a}\hat{a}$~\cite{GZ,Fernandez}.
In this work we deal with the parameter ranges in which such Kerr
nonlinearity can be neglected, and hence classical optical
bistability does not appear. Moreover, we consider the regime in
which the mechanical effects of photon-atom interactions are
relevant. In this range, the nonlinearity in the dynamics of
Hamiltonian~(\ref{jcdisp}) is solely due to the coupling between the
cavity field and the atomic quantum center-of-mass variables.

\subsection{Many-body Hamiltonian}

We now introduce the system which we investigate in this paper,
namely a gas of $N$ fully polarized fermionic atoms interacting
dispersively with the mode of the resonator. Denoting by
$\hat{\Psi}(x)$ the atomic field operators obeying Fermi
commutation rules, the rescaled Hamiltonian in second quantization reads
\begin{eqnarray}\label{2-quant} && H -\Delta_c\hat{a}^\dagger\hat{a}-i\eta\left(\hat{a}-\hat{a}^\dagger\right)\\
&&+\!\int{\rm
d}x\!\left(\!-\hat{\Psi}^\dagger(x)\frac{d^2}{dx^2}\hat{\Psi}(x)
\!+\!U_0\cos^2(
x)\hat{a}^\dagger\hat{\Psi}^\dagger(x)\hat{\Psi}(x)\hat{a}\!\right)\!,\nonumber
\end{eqnarray} where we have taken care of the proper ordering between
atomic and field operators, which is here normal (see also~\cite{jonas2}).

Let us now consider that the number of photons is fixed to the value $n$. Then, after tracing out the
photonic degrees of freedom from Hamiltonian~(\ref{2-quant}), the
resulting Hamiltonian describes the motion of $N$ Fermions in the
optical lattice of the cavity field, whose depth is proportional
to the number of photons $n$. Hence, the degree of localization of the atoms at the minima of the potential wells depends on $n$, and so will the Wannier functions in the tight-binding limit. We now assume that the atoms are well
localized at the minima of the potential, and expand the atomic field operators in the tight-binding
limit~\cite{mermin} in the corresponding Wannier functions for the lowest Bloch band,
\begin{equation}\label{Wannier}
\hat\Psi(x)=\sum_{i=1}^K\hat{f}_iw_{\hat{n}}(x-x_i)\end{equation}
where $\hat{f}_i$ is the annihilation operator of a fermion at
site $i$, $w_{\hat{n}}(x-x_i)$ is the Wannier function localized
at site $i$, and $K$ is the number of lattice sites. The subscript
$\hat{n}$ indicates the operator valued dependence of the Wannier
functions on the number of photons
$\hat{n}=\hat{a}^{\dagger}\hat{a}$. Such an expansion, considering only the lowest band, is known to be justified for large or moderate lattize depths, which we will discuss in more detail later on. Using the Wannier expansion
within the single-band approximation, Eq.~(\ref{Wannier}), we
obtain from Eq.~(\ref{2-quant}) the Hamiltonian \begin{equation}
\label{BH-Fermi} \begin{array}{lll}
\mathcal{H} & = &\displaystyle{\sum_{i,j=1}^K\left[E_{ij}(\hat{n})+U_0\left(\hat{a}^\dagger J_{ij}(\hat{n})\hat{a}\right)\right]\hat{f}_i^\dagger\hat{f}_j}\\
\\
& &
-\Delta_c\hat{a}^\dagger\hat{a}-i\eta\left(\hat{a}-\hat{a}^\dagger
\right) \end{array} \end{equation} where
\begin{equation}\label{para} \begin{array}{l}
\displaystyle{E_{ij}(\hat{n})=\int dx\,w_{\hat{n}}^*(x-x_i)\left(-\frac{d^2}{dx^2}\right)w_{\hat{n}}(x-x_{j})},\\ \\
\displaystyle{J_{ij}(\hat{n})=\int dx\,w_{\hat{n}}^*(x-x_i)\cos^2(x)w_{\hat{n}}(x-x_{j})},\\ \\
\end{array} \end{equation} are the coupling parameters which
depend on the number of photons of the cavity field. In the
tight-binding approximation we keep only on-site and
nearest-neighbour interactions in Eq.~(\ref{BH-Fermi}), and denote
the relevant coupling parameters by $E\equiv E_{ii}$, $J\equiv
J_{ii}$, $E_1\equiv E_{i,i+1}$, and $J_1\equiv J_{i,i+1}$. This is legitimate in the single band regime. We impose periodic boundary conditions and use the representation
$\hat{c}_k=\sum_j\hat{f}_j{\rm e}^{{\rm i}k j}$, where $\hat{c}_k$
($\hat{c}_k^{\dagger}$) annihilates (creates) a fermion at the
rescaled quasi-momentum $k$, with $k$ in the interval $[-1,1]$ and
step $\delta k=2/K$. In this representation, the number operator
$\hat{N}=\sum_j\hat{f}_j^{\dagger}\hat{f}_j$ and the hopping
operator $\hat{B}=\sum_j\hat{f}^{\dagger}_j\hat{f}_{j+1}+{\rm H.c.}$
get the diagonal forms
\begin{equation}\label{nb} \begin{array}{l}
\hat{N}=\sum_k\hat{c}_k^\dagger\hat{c}_k, \\ \\
\hat{B}=2\sum_k\cos(k\pi)\hat{c}_k^\dagger\hat{c}_k, \end{array}
\end{equation} Hence, the new Hamiltonian takes the form
\begin{eqnarray}\label{ham1}
&&\mathcal{H}=\!-\Delta_c\hat{a}^\dagger\hat{a}-{\rm i}
\eta\!\left(\hat{a}-\hat{a}^\dagger
\right)\!+\mathcal{H}_1(\hat{n})\!+\mathcal{H}_2(\hat{n})\hat{a}^\dagger\hat{a}
\end{eqnarray} with
\begin{equation}
\begin{array}{l}
\displaystyle{\mathcal{H}_1(\hat{n})=\sum_{k}\Big[E+2E_1\cos(k\pi)\Big]\hat{c}_k^\dagger\hat{c}_k,}
\\ \\
\displaystyle{\mathcal{H}_2(\hat{n})=U_0\sum_{k}\Big[J(\hat n - 1 )
+J_1(\hat n - 1 )2\cos(k\pi)\Big]\hat{c}_k^\dagger\hat{c}_k.}
\label{H:2}
\end{array}
\end{equation}
Here, we used the relation ${\mathcal
F}(\hat{n})\hat{a}=\hat{a}{\mathcal F}(\hat{n}-1)$ for any
function $\mathcal F(z)$ which is analytic in the scalar variable
$z$.

\subsection{Nonlinearities in the coupling parameters}

Equation~(\ref{ham1}) describes dynamics, which couple in a nonlinear way cavity and atomic degrees of freedom. In particular, the coupling parameters $E$, $E_1$, $J$, $J_1$ depend on the number of photons, since they are integrals of Wannier functions, which themselves depend on the intensity of the cavity field. In order to better understand the character of this dependence on the intracavity photon number we evaluate their explicit form by replacing the Wannier functions in Eqs.~(\ref{para}) by Gaussians functions,
\begin{equation}
w_{\hat{n}}(x-x_i)\approx
w_{\hat{n}}^G(x-x_i)=\frac{1}{\sqrt[4]{\pi\sigma^2}}\mathrm{e}^{-\frac{(x-x_i)^2}{2\sigma^2}},
\end{equation}
where $\sigma=|V|^{-1/2}$ and $V=U_0n$ is the potential depth at a
fixed number of photons $n$. Moreover, we modify the Gaussian
functions by imposing the condition $\int
dx\,w_{\hat{n}}^G(x-x_i)w_{\hat{n}}^G(x-x_j)=\delta_{ij}$, which
allows us to avoid non-physical contributions. This ansatz is
usually valid within the tight-binding approximation as will be further discussed in Sec.~\ref{ssec3d} and has been checked in~\cite{jonas2}. Using these modified Gaussian functions we
solve the integrals in Eq. (\ref{para}) and obtain
\begin{equation}\label{gausspara}
\begin{array}{l}
\displaystyle{E=\frac{1}{y}},\\ \\
\displaystyle{J=\frac{1}{2}\left(1-s\mathrm{e}^{-y}\right)},\\ \\
\displaystyle{E_1=-\frac{1}{2y^2}\mathrm{e}^{-\frac{\pi^2}{4y}}\left(2y+\pi^2\right)},\\ \\
\displaystyle{J_1=s\frac{1}{2}\mathrm{e}^{-\frac{\pi^2}{4y}}\mathrm{e}^{-y}},
\end{array}
\end{equation}
where we have introduced $y=\sigma^2=|V|^{-1/2}$ and
$s=\mathrm{sign}(\Delta_a)$. Relaxing the orthogonality condition one would find that the amplitude $|J_1|$ is different for red or blue detuning. The sign dependence $s$ in $J$ and $J_1$ arisses from the fact that the Wannier functions are centered at different positions in the two cases. From these expressions we find $|J_1|<
18J$ within the validity of the tight-binding approximation, which
according to Ref.~\cite{jonas2} is given by $y<0.5-1$. Moreover, the
contribution from $\ell$-neighbours can be estimated to be
$|J_{\ell}/J_1|=\exp\big[-({\ell}^2-1)\frac{\pi^2}{4y}\big]\ll 1$,
which justifies neglecting next nearest-neighbour terms and so forth
in Eqs.~(\ref{ham1}).

\section{Dynamics and steady state of the cavity field} \label{Sec:III}

In this section we study analytically and numerically the dynamics
and steady state of the cavity field, focussing on instabilities
due to the interplay between field and matter wave fluctuations at
a fixed number of atoms $N$. We consider the Heisenberg equations
of motion for the field operators, 
\begin{eqnarray}
\dot{\hat{a}}
& = &
-i[\hat{a},\mathcal{H}_1(\hat{n})]-i[\hat{a},\mathcal{H}_2(\hat{n})]\hat{n} \nonumber \label{heis:1}\\
& & -i\mathcal{H}_2(\hat{n})\hat{a}+i\Delta_c\hat{a}-\kappa\hat{a}+\eta+\hat{\xi}_a,\\ \nonumber \\
\dot{\hat{n}} & = & \eta\left(\hat{a}+\hat{a}^\dagger
\right)-2\kappa\hat{n}+\hat{\xi}_n, \label{heis:2}\end{eqnarray} 
and similarly for $\hat{a}^\dagger$. These are
coupled to the equations for the atomic field operators, and where
$\hat{\xi}_i$ are the quantum noise operators, such that $\langle
\hat\xi_i(t)\rangle =0$ and $\langle
\hat\xi_i(t)\hat\xi_i^{\dagger}(t')\rangle =2\kappa\delta
(t-t^{\prime})$ for $i=a,\,n$. Such fluctuations vary on time scales which are
faster than $1/\kappa$, and will be neglected when taking the mean
values on the coarse-grained time scales, at which we analyze the
dynamics in what follows. The numerics are carried out by first evaluating the commutators and then replacing the operators by $c$-numbers, $\hat{a}\rightarrow\alpha$, $\hat{a}^\dagger\rightarrow\alpha^*$ and $\hat{n}\rightarrow\bar{n}=|\alpha|^2$, and then solving the differential equations using the regular Runge-Kutta method. As initial conditions we thus use $\bar{n}(t=0)=n_0$, $\bar{a}=\sqrt{n_0}$ and $\bar{a}^*=\sqrt{n_0}$ for some real $n_0$. Note that the commutators are carried out exactly, {\it i.e.} $[\hat{a},g(\hat{n})]=\hat{a}\left(g(\hat{n})-g(\hat{n}-1)\right)$, and the $c$-number replacement is done afterwards. Truncating the system dynamics to only three equations and dropping the operator structure should be justified for the open system we consider.

\subsection{Intracavity field intensity}

The intracavity intensity is found from the mean value of photons,
\begin{equation} \bar{n}(t)=\langle \hat n(t)\rangle.
\label{mean:n} \end{equation} This average is taken over field and
atomic degrees of freedom, assuming that the cavity field variables
relax to a stationary value on a faster time scale than the time
scale of the evolution of the atomic variables, hence for $\kappa\gg
k_BT/\hbar$, with $\kappa$ the cavity decay rate, $k_B$
Boltzmann's constant and $T$ the temperature. In this regime, a
closed expression for the photon number operator can be found in the
limit in which the number of photons is sufficiently large (In fact, the analytic expression gives to a good approximation also the steady state solution for weak fields.). In this
regime we approximate
\begin{equation}\label{derapp}
{\mathcal F}_2(\hat{n}-1)\approx {\mathcal
F}_2(\hat{n})-\partial{\mathcal F}_2^{\prime}(\hat{n}),
\end{equation}
where $\mathcal F$ is any analytical function and $\partial$
indicates derivative with respect to photon number, and from
Eq.~(\ref{H:2}) we find
\begin{equation} \mathcal{H}_2(\hat{n})\approx
U_0\sum_{k}\Big[(J-\partial J)+(J_1-\partial
J_1)2\cos(k\pi)\Big]\hat{c}_k^\dagger\hat{c}_k.
\end{equation}
Using similar relations for the commutators in Eq.~(\ref{heis:2}),
we find an analytical expression that the mean photon number
has to fulfill, \begin{equation}
\bar{n}=\frac{\eta^2}{\kappa^2+\big(\Delta_c-\xi(\bar{n})\big)^2}
\end{equation} where
\begin{equation} \label{zeta:n} \xi(\hat
n)=\mathcal{H}_2+\partial\mathcal{H}_2+\partial\mathcal{H}_1+\partial\mathcal{H}_2\hat{n}.
\end{equation}
In order to evaluate Eq.~(\ref{mean:n}) we assume that the atomic
state is the ground state of the potential, whose depth is
determined by the time-dependent number of photons $\bar{n}(t)$. The
ground state is then evaluated assuming that the tight-binding and
the single-band approximations are valid during the evolution of the
field. This treatment is hence valid in specific parameter regimes,
namely when the rate of variation of the photon number, giving the
height of the cavity potential, remains smaller than the energy gap
between the lowest and the first band, and when the mean number of
photons $\bar{n}$ warrants sufficient localization of the atoms,
according to the relation $$y=\frac{1}{\sqrt{|U_0|\bar n}}\ll 1.$$
We have numerically checked that for $y<0.5$ (which
coincides with the tight-binding and Gaussian approximations), the
first band gap is larger than the recoil energy. Thus, remaining in the validity regimes for the tight-binding and Gaussian approximations ($y<0.5$), also warrants no non-adiabatic coupling of excited bands. Further, the large band gap makes the system resistant to nonzero temperature excitations.  

\subsection{Many-body atomic state}

In order to calculate any quantities we need to specify the particular state of the atoms. In the regime in
which we can assume that the cavity field variables relax to a
stationary value on a faster time scale than the time scale of the
evolution of the atomic variables, we can eliminate the photon
variables from the atomic Heisenberg equations. In doing this we
follow the procedure discussed in~\cite{jonas2}, now applied to
fermionic atoms. Hence finding that the atomic dynamics are
determined by an effective Hamiltonian of the form
\begin{equation}\label{asham} \mathcal{H}_{\rm
eff}=E\hat{N}+f(\hat{N})+\left[E_1+\frac{U_0\eta^2J_1}{\kappa^2+\zeta^2(\hat{N})}\right]\hat{B},
\end{equation} where \begin{equation} \begin{array}{ll}
\displaystyle{f(\hat{N})=\frac{\eta^2}{\kappa}\arctan\left[\frac{\zeta(\hat{N})}{\kappa}\right]},\\ \\
\displaystyle{\zeta(\hat{N})=\Delta_c-U_0J_0\hat{N}} \end{array}
\end{equation} and the coupling terms given in (\ref{gausspara}).
This Hamiltonian is derived at leading order in an expansion in
$1/N$, assuming a scaling such that the number of lattice sites $K$
is proportional to the number of atoms $N$ where we assume a fixed
number of atoms $N$. The coefficients of this Hamiltonian depend on
the atomic variables through the nonlinear coupling with the cavity
field~\cite{jonas1,jonas2}. When they are independent of the atomic
variables, for instance in the case of Fermi gases in deep optical
lattices in free space, the ground state of this Hamiltonian is the
usual Fermi sea, where the value $\sum_{i=1}^N|k_i|$ is minimized.
Thus, the many body ground state is found by filling with one atom
all states with quasi-momentum $k$ up to the Fermi quasi-momentum
$k_F$, assuming that $k_F$ is
smaller than the edge of the Brillouin zone.
This simple procedure cannot be applied when the nonlinearity in the
Hamiltonian gives dispersion curves that may exhibit local minima,
which can be at values of $k$ different from $k=0$ (at the center of
the Brillouin zone). Finding the distribution that gives the
smallest energy is in general a linear programming (LP)
optimization problem with two constrains: (i) at most one particle
can occupy each site and (ii) the total number of particles equals
$N$. However, since the number of particles per site is a discrete
variable, $n_i=0,1$, standard LP techniques are not applicable. We
use instead a variational approach where we assume that the usual
Fermi sea is the ground state, and check its stability against
perturbations. Such check is made by randomly extracting a single
fermion from the Fermi sea and putting it in a random state
outside the sea. We then compare the energies, calculated using
$\mathcal{H}_{\rm eff}$ in Eq.~(\ref{asham}), for the Fermi sea and the
perturbed state. Repeating this procedure sufficiently many times,
we find that the energy of the Fermi sea (according to our ansatz) is usually lower than the
perturbed one. In some cases, however, the perturbed state is
energetically favorable. On the other hand, the difference in energy
between the two states is extremely small for the parameter regimes
of interest and hence does not affect the general structure of our
results. We have also compared the energy of a fully random
distribution with the one of the Fermi sea and in all examples we considered found
the usual Fermi sea to have a lower energy. Still, these situations
are potentially of great interest, since they may indicate novel
quantum phase transitions due to the change of geometry, or even
topology of the Fermi surface, similar to what happens in graphene
(cf. \cite{grap1,grap2}). These questions go beyond the focus of the
present paper, and will be studied elsewhere.

\subsection{Quantum optical bistability}

We now evaluate Eq.~(\ref{mean:n}) assuming that the atoms are in
the state described by the usual Fermi sea, and find a nonlinear
equation for $\bar{n}$, which may exhibit multiple solutions. We
remind that such nonlinearity is not due to the nonlinear coupling
between atomic dipoles and fields, but instead originates from the
quantum fluctuations of the atoms at the minima of the confining
potential. In order to highlight this behaviour, we express
$\bar{n}=1/U_0y^2$, using the relation between the depth of the
potential and the width of the Gaussian functions. From
Eq.~(\ref{mean:n}), using the approximation
$[\hat{a},g(\hat{n})]\approx\frac{\partial
g(\hat{n})}{\partial\hat{n}}\hat{a}\equiv\partial g(\hat{n})\hat{a}$
in Eq.~(\ref{zeta:n}), we find the nonlinear equation which
determines $y$,
\begin{eqnarray}\label{ssres}|U_0|\eta^2y^2-\kappa \left\{\Delta_c-\frac{U_0N}{2}\left(f_1(y)+\tilde{B}{\rm
e}^{-\frac{\pi^2}{4y}}f_2(y)\right)\right\}^2\nonumber\\
\label{secondsol} \end{eqnarray} where
$\tilde{B}=2\sum_{i=1}^N\cos(k_i\pi)/N$ is the mean value of the
operator $\langle B\rangle/N$, taken on the Fermi sea of atoms, and
\begin{eqnarray} &&f_1(y)=1+s y-s {\rm
e}^{-y}(1+y/2)\\
&&f_2(y)=s {\rm
e}^{-y}\!\left(\!1+\frac{y}{2}-\frac{\pi^2}{8y}\right)\!+\frac{\pi^4}{8y}-\frac{6\pi^2}{8}-y.
\end{eqnarray} Here we neglected small terms such as the
derivative of the $E_1,J_1$-terms, which indeed has been verified to only give minimal corrections. 

\begin{figure}[ht] \begin{center}
\includegraphics[width=8cm]{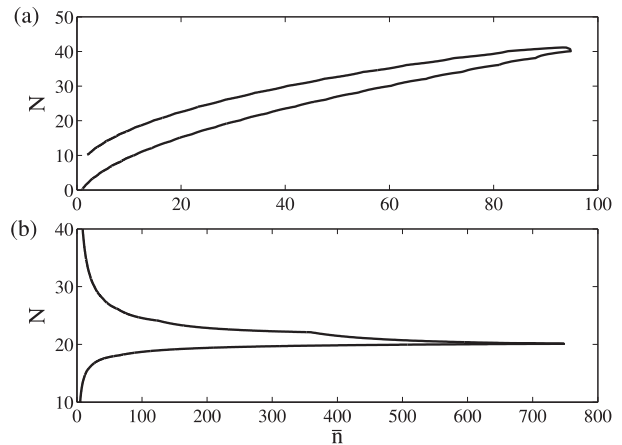}
\caption{Set
of parameters $(\bar{n},N)$ for which Eq.~(\ref{secondsol}) is
satisfied. The dimensionless parameters are $\kappa=1$, and (a):
$U_0=10$, $\eta=10$, $\Delta_c=10$ and (b): $U_0=-1$, $\eta=30$,
$\Delta_c=-20$. In both plots we assumed $K=50$ sites of the cavity
potential. The curves are shown for the values of $(\bar{n},N)$ such
that $y<0.5$, when the tight-binding approximation is valid. }
\label{fig1}
\end{center}
\end{figure}

Figure~\ref{fig1} displays the set of values $(\bar{n},N)$ for which
Eq.~(\ref{secondsol}) is satisfied. The parameter regime is
restricted to $y<0.5$, where our treatment is valid. In particular,
in Fig.~\ref{fig1} (a) the detuning $\Delta_a>0$ and the atoms
are trapped at the nodes of the cavity potential, where for maximum
localization, $y\to 0$, the coupling with the resonator vanishes.
Nonlinearity here arises due to the finite size of the fluctuations.
In Fig.~\ref{fig1}(b), instead, $\Delta_a<0$, and the atoms are
confined at the antinodes, where the coupling to the field is
maximum. For large atom numbers the effect of small position
fluctuations at these points gives a small correction to the cavity
intensity, and the system shows less sensitivity on the external
parameters, see also~\cite{jonas1,jonas2}.

\begin{figure}[ht] \begin{center}
\includegraphics[width=8cm]{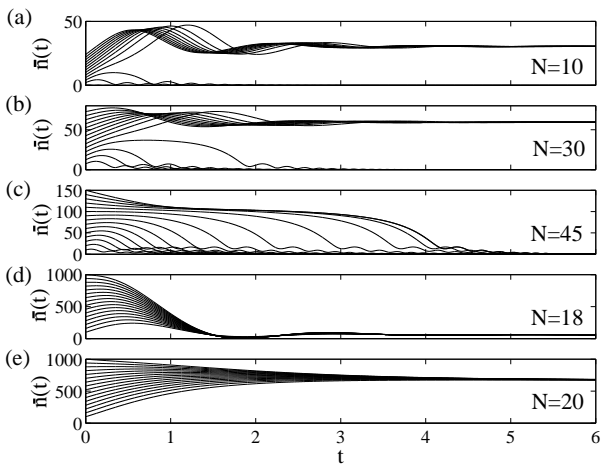}
\caption{Mean number of photons $\bar{n}$ as a function of time,
obtained by numerically integrating the Heisenberg equations in
Eqs.~(\ref{heis:1})-(\ref{heis:2}) for different initial values
$n(0)$ and for a fixed number of fermionic atoms $N$. In (a)-(c) the
parameters correspond to the ones of Fig.~\ref{fig1}(a), in (d),(e)
to the ones of Fig.~\ref{fig1}(b). In the calculations we assumed
that the atoms are in the ground state of the cavity potential, and
that the tight-binding regime and the single-band approximation are
valid. This is fulfilled for $\bar{n}>0.4 $ in (a)-(c), and for
$\bar{n}>4$ in (d),(e). Once the curves reach below these limiting
values, they lie outside the validity regime of our approximations.
} \label{fig2} \end{center} \end{figure}

In order to check the stability of the solutions, we numerically
solve the Heisenberg equations in Eqs.~(\ref{heis:1})-(\ref{heis:2})
for various initial values of $\bar{n}$ at a fixed number of atoms
$N$, as previously explained. Figure~\ref{fig2} displays the evolution of some different
values of the mean number of photons $\bar{n}(t)$. In particular,
Figs.~\ref{fig2}(a)-(c) have been evaluated in the same parameter
regime of Fig.~\ref{fig1}(a) and for a particular choice of the
number of atoms $N$. In Fig.~\ref{fig2} (a) and (b) it is visible
that the mean value exhibits two possible asymptotic solutions,
which depend on the initial value, while in Fig.~\ref{fig2}~(c)
there is only one asymptotic value of $\bar{n}$. Comparing with the
results in Fig.~\ref{fig1}(a), we find that only the solutions of
the lower branch of the curve $(\bar{n},N)$ in that figure are
recovered, while the upper branch is unstable. The additional
solutions with small asymptotic $\bar{n}$ in fig.~\ref{fig2} (a) and
(b) correspond to $y>0.5$ and are thus beyond the validity regime of
our approximations and hence not visible in Fig.~\ref{fig1} (a). The
lower two plots, Figs.~\ref{fig2}(d) and~(e), should be compared with
Fig.~\ref{fig1}(b). Here, one can observe that a small change in atomic
number, $N=20$ or $N=18$, results in intracavity field intensities
differing by a factor of about 100. We again pointed out that
by calculating the Heisenberg equations of motion we use the exact
relation
$[\hat{a},g(\hat{n})]=\left(g(\hat{n})-g(\hat{n}-1)\right)\hat{a}$,
and do not impose any approximations involving derivatives as in
(\ref{derapp}). Still, the asymptotic solutions of the numerical
integration agree well with the ones obtained from
(\ref{secondsol}), which verifies the trustworthiness of the steady state results \ref{secondsol}).

\begin{figure}[ht] \begin{center}
\includegraphics[width=8cm]{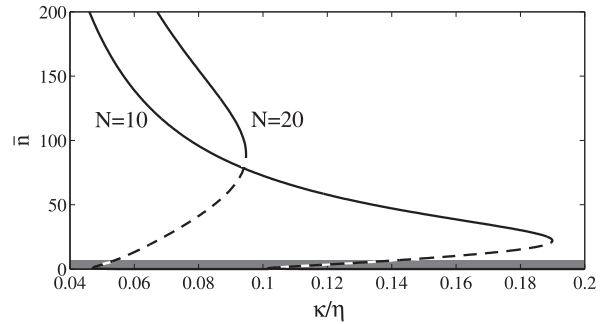}
\caption{Mean number of photons $\bar{n}$, evaluated in
Eq.(\ref{secondsol}) for a fixed number of atoms $N=10,20$, as
function of the pumping $1/\eta$ (in units of $1/\kappa$). The solid
(dashed) line indicates the stable (unstable) solutions. The
dimensionless parameters are $U_0=0.62$ and $\Delta_c=5$, and the
sites of the cavity potential are $K=50$. The region shaded in grey
corresponds to the values $\bar{n}<7$ where the tight-binding and
single-band approximations are not valid.  } \label{fig3}
\end{center} \end{figure}

We finally study the instability of the mean number of photons $\bar n$,
and hence the cavity potential depth, as the pumping amplitude is varied. Typical examples are shown in Fig.~\ref{fig3}, where the
solutions for $\bar{n},N$ of Eq.~(\ref{secondsol}) are plotted as a
function of the pump amplitude $\eta$. The dashed middle branch
represents an unstable solution of the equations. Note that the
lower branch is almost identical to $\bar{n}=0$. We see that $\bar
n$ exhibits jumps at critical values of the pump $\eta$,
corresponding to an instability due to the nonlinear behaviour, to
values in which our approximations are invalid. Such
regime is in the grey-shaded zone. Here, our treatment is invalid
and we cannot make definite predictions. We conjecture that at
these points the fermions are populating higher Bloch bands or
forming a Fermi liquid. The study of this transition will be
explored in detail in future works.

\subsection{Validity of approximations and typical experimental parameters}\label{ssec3d}
Various approximations have been imposed throughout the paper and it is in order to summarize them 
and discuss their reliabilities. The very first assumption of our model system is the adiabatic elimination of 
the excited atomic levels and neglecting spontaneous emission. The justification of leaving out spontaneous emission, 
and at the same time taking into account for the cavity spectral line width and the single mode assumption, 
was discussed in detail in \cite{jonas2}. There we found effective atom-field couplings $g_0\sim2\pi\times100$ MHz 
or $g_0\sim2\pi\times700$ MHz using, respectively, the cavity decay rates $\kappa$ of \cite{Esslinger} 
or \cite{Reichel}. These are presently slightly outside the regime of experimental reach. This may, 
however, be circumvented by increasing the total number of atoms $N$ \cite{jonas2}. 
To assure the elimination of the excited atomic state we should have $\Delta_a\sim10g_0\sqrt{\bar{n}}$,
 where the average number of photons $\bar{n}$ is typically 100. This suggests $U_0<g_0$ and in our numerical 
analysis $U_0$ is of the same order as $\kappa$, while for realistic parameters one has $g_0>\kappa$ and thus ensure 
that the upper atomic state can be eliminated.

We further imply the Gaussian, single band and tight-binding 
approximations in order to derive the many-body Hamiltonian (\ref{ham1}). 
The validity of these approximations is thoroughly analyzed in \cite{jonas2} where it is 
found that they all break down for $y>1$, or equivalently $|V|<1$. In this work we consider
 $y<0.5$ (or otherwise explicitly pointed out) to meet this constrain. Especially, in the instability 
of field intensities as the pump amplitude is varied, shown in Fig.~\ref{fig3}, are these approximations assumed 
to be violated. However, instability is still predicted, jumping from a strong cavity field to a weak one even 
if the true full state of the system cannot be predicted. Related is the conjecture of the Fermi sea state of the atoms. 
This requires an adiabatic change of any system parameters, but we have verified that our results are reproducible 
also for deviations from the Fermi sea state and therefore not restricted to the fully adiabatic limit. 
Consequently, the system is also robust against small but nonzero temperatures. 

In solving for the dynamics, the set of Heisenberg equations of motion (\ref{heis:1}) 
and (\ref{heis:2}) is truncated to contain the equations for $\hat{a}$, $\hat{a}^\dagger$ and $\hat{n}$, 
which is believed to be justified in the regime of fairly large cavity decay considered here. The analytical 
steady state result for the field amplitude uses $[\hat{a},g(\hat{n})]\approx\partial g(\hat{n})\hat{a}$ and also 
neglects small terms in the final expression (\ref{ssres}). This is supported since the results of the steady state 
numerical calculations of the Heisenberg equations coincide well with ones predicted by (\ref{ssres}), even for small 
field amplitudes $\bar{n}<10$.

To get an idea about characteristic experimental scales and quantities, 
we consider potassium atoms $^{40}\mathrm{K}$ and a wave length $\lambda=800$ nm, 
giving the recoil frequency $\omega_r\approx50$ kHz. Thus, $t=1$ corresponds to an unscalled 
time $\tilde{t}=20\,\,\mu$s, and estimating the critical temperature as $T=\hbar\omega_r/k_B$ 
one finds $T\sim0.4\,\,\mu\mathrm{K}$. Note that cavity decay rates used in this paper are of order of the recoil 
frequency, i.e. realistic but  rather small. 
However, we have verified that $\kappa$ can be increased, while $U_0$, $\eta$ 
and $\Delta_c$ are scaled accordingly, and our results still remains.  

\section{Discussion and conclusions}\label{Sec:V}

We have reported on instability in the intracavity mean photon
number, which are driven by the pump intensity and by the atomic
density. Differing from usual optical bistability, these nonlinear
effects arise in the regime in which the atomic polarization is
linear in the intensity of the cavity field. In fact, such
instability is due to nonlinear coupling between the atomic motion
and the cavity field, whose dynamics and steady state depend in a
complex way on various parameters, such as pump amplitude,
frequency, and the atomic density distribution. The system we here
consider is a gas of ultracold Fermi atoms, and the nonlinearity we
observe is exclusively due to the coupling of the atoms with the
resonator, whereby particle-particle collisions are neglected due to
the Pauli principle. The atomic density distribution hence enters in
determining the strength of the coupling of the atomic fermions to
the cavity field, accordingly the height of the cavity potential
which finally, closing this nonlinear circle, determines the atomic
density distribution itself.

The dependence of the cavity field dynamics on the atomic density
is indeed peculiar: At different atomic densities, different
behaviours and instabilities of the cavity field are observed. One
could hence conjecture that this system may also exhibit
fluctuations and instabilities in the mean number of atoms, in the
regime in which this is not fixed, as it could be observed when
the gas is put in contact with a particle reservoir.

Further interesting outlooks emerge when considering the situation
at larger intracavity intensities, in which the atomic polarization
is nonlinear in the field. In this case, this nonlinearity, which
corresponds to the one of dispersive optical bistability, adds to
the one of the coupling to the motion, and gives rise to a
competition between cavity potentials. In fact, at lower intensities
the potential is of the form $\cos^2(\hat{X})$, while at larger
fields it is a superposition of $\cos^2(\hat{X})$ and
$\cos^4(\hat{X})$. Hence, instabilities in the atomic ground states
may emerge, which are expected to give rise to novel behaviours.
Such rich scenario is indeed exciting and requires different
approaches to the study of this problem, which is under current investigation \cite{jonasjani}.

\begin{acknowledgements}
We thank Jakob Reichel, Stefan Ritter,
and Mirta Rodriguez for helpful discussions. We acknowledge
support of the Integrated Project "SCALA" of the European
Commission (Contract No.\ 015714), the grants of the Spanish Ministerio de Educaci\'on y
Ciencia (FIS 2005-04627; Consolider Ingenio 2010 "QOIT";
Ramon-y-Cajal; QNLP, FIS2007-66944), and the ESF-MEC Program "EUROQUAM" ("CMMC", "Fermix"). J.L. acknowledges support
from the Swedish government/Vetenskapsr{\aa}det.
\end{acknowledgements}

\end{document}